\documentclass[journal,twoside]{IEEEtranTCOM}
\usepackage{amsopn, amsmath, fancybox, epsfig, amssymb, color, bm, dsfont , latexsym, graphicx,
accents, subfigure, algorithm, algpseudocode, multirow,  pifont, wasysym, stmaryrd}

\newcommand{\BSGN}[1]{\text{bsgn}#1}

\newcommand{\squeezeup}{\vspace{-0.4em}}

\begin{document}

\title{Adaptive Group Shuffled Decoding for LDPC Codes}

\author{Tofar~C.-Y.~Chang,~\IEEEmembership{Member,~IEEE}~and~Yu~T.~Su,~\IEEEmembership{Senior Member,~IEEE}}

\maketitle
\begin{abstract}
We propose new grouping methods for group shuffled (GS) decoding of both regular and irregular
low-density parity check cods. These methods are independent of the check-to-variable message
formula used. Integer-valued metrics for measuring the reliability of each tentative variable
node (VN) decision and the associated likelihood of being corrected are developed. The metrics
are used to determine the VN updating priority so the grouping may vary in each iteration. We
estimate the computation complexity needed to adaptively regroup VNs. Numerical results show
that our GS algorithms improve the performance of some existing GS belief-propagation decoders.
\end{abstract}


\squeezeup
\section{Introduction}
Conventional belief-propagation (BP) based algorithms for decoding low-density parity check (LDPC) codes
update all variable nodes (VNs) and check nodes (CNs) in parallel \cite{LDPC}. The group shuffled (GS)
BP algorithm \cite{SBP2} partitions the VNs into groups and performs group-wise parallel decoding serially
which in effect divides a decoding iteration into several sub-iterations. A GSBP decoder can thus pass
newly updated messages obtained in a sub-iteration to the neighboring groups in the ensuing sub-iterations
and achieve faster convergence with reduced parallel decoding complexity. The GSBP algorithm generalizes
the original shuffled BP algorithm \cite{SBP2} or the column-based layer BP decoder \cite{LBP} by allowing
a group to contain more than one VN. Some improved GSBP algorithms have been proposed to further improve
the decoding performance \cite{GSBP1}--\cite{GSBP3}. A variation of the GSBP approach which greedily selects
the `best' edge(s) for updating is the class of informed dynamic scheduling based BP decoders \cite{IDS1},
\cite{IDS3}. However, the greedy search requires high computing complexity and some VNs may never or seldom
be updated.

One of the basic ideas of our GS decoding schedule is to prioritize the updates of the VNs which most probably
have erroneous tentative decisions and are likely to be corrected. Early-updating such VNs enables us to invert
an incorrect tentative local decision, avert potential error propagation, and strengthen the reliability of the
passed messages. This concept is an extension of those inspire the algorithms presented in \cite{IDS1},\cite{IDS3}.
We not only determine the set of VNs with the most unreliable tentative decision but also evaluate
and compare the impact/benefit of updating these VNs. In this letter, we develop a set of simple binary/integer
based rules that dynamically re-group VNs in each iteration for GS decoding. As will be seen, the improved
performance comes at the expense of extra integer and binary operations (against the conventional GSBP decoders)
and provides additional complexity-performance trade-offs in designing a BP decoding schedule.

The rest of this letter is organized as follows. In Section \ref{section:GSBP}, we define the basic system
parameters and give a brief review of the standard GSBP decoding algorithm. In Section \ref{section:Proposal},
we introduce our adaptive GSBP (AGSBP) algorithm and two adaptive grouping approaches. Numerical results are
presented in Section \ref{section:results}, and the complexity of the AGSBP algorithms are also analyzed in the
same section. Finally, conclusion remarks are drawn in Section \ref{section:conclusion}.
\squeezeup
\section{Group Shuffled BP Decoding}
\label{section:GSBP} A binary ($N$, $K$) LDPC code $\mathcal{C}$ is a linear block code of rate $R= K/N$ described
by an $M \times N$ parity check matrix $\bm{H}$ which has $d^v(n)$ ones in the $n$th column and $d^c(m)$ ones in
the $m$th row. $\bm{H}$ can be viewed as a bipartite graph with $N$ VNs corresponding to the encoded bits and $M$
CNs corresponding to the parity-check functions represented by the rows of $\bm{H}$. The conventional GSBP algorithm
\cite{SBP2} divides the VNs into $G$ groups of equal size $N/G=N_G$ according to their natural order, i.e., if we
define $\mathcal{G}_i=\{n|i\cdot N_G\leq n<(i+1)\cdot N_G-1\}$ where $i=0,1,\ldots,N-1$, then VN $v_n$ belongs to
the $i$th VN group if $n \in \mathcal{G}_i$. In each GSBP decoding iteration, groups are sequentially processed and
the VNs belonging to the same group are updated in parallel.

A binary codeword $\bm{u}=(u_0,u_1,\cdots,u_{N-1})$ is BPSK-modulated and transmitted over an zero-mean AWGN channel
with noise variance $\sigma^2$. The corresponding received and binary decoded sequences are denoted by $\bm{r}=(r_0,r_1,
\cdots,r_{N-1})$ and $\hat{\bm{u}}=(\hat{u}_0,\hat{u}_1,\cdots,\hat{u}_{N-1})$. We define $m^v_{n\rightarrow m}$ as
the variable-to-check (V2C) message from the $n$th VN $v_n$ to the $m$th CN $c_m$ and $m^c_{m\rightarrow n}$ as the
check-to-variable (C2V) message from $c_m$ to $v_n$. Let $\mathcal{N}$($m$) be the index set of VNs which are connected
to $c_m$ and $\mathcal{M}$($n$) be that of CNs connected to $v_n$ in the code graph. $\mathcal{N}(m)\backslash n$ is
the set $\mathcal{N}$($m$) with $n$ excluded; $\mathcal{M}(n)\backslash m$ is similarly defined. We further define the
sign function $\text{sgn}(x)=1$ if $x>0$, $\text{sgn}(x)=-1$ if $x<0$, and takes the value $1$ or $-1$ equally likely
if $x=0$. We assume that the log-domain BP decoding is used. When processing the $i$th VN group, the C2V messages
$m^c_{m\rightarrow n}$, $\forall m\in\mathcal{M}(n),n\in \mathcal{G}_i$ are computed by
\begin{eqnarray}\label{eqn:C2V}
m^c_{m\rightarrow n}=\prod_{n'\in\mathcal{N}(m)\setminus n}\alpha_{n'\rightarrow m}\cdot\phi\left(\sum_{n'\in\mathcal{N}(m)\setminus n}
\phi(\beta_{n'\rightarrow m})\right)
\end{eqnarray}
where $\phi(x)=-\log[\tanh(x/2)]$, $\alpha_{n'\rightarrow m}=\text{sgn}(m^v_{n'\rightarrow m})$,
and $\beta_{n'\rightarrow m}=|m^v_{n'\rightarrow m}|$. The V2C messages $m^v_{n\rightarrow m},
\forall~m\in\mathcal{M}(n),~n\in \mathcal{G}_i$, are updated via
\begin{eqnarray}\label{eqn:V2C}
    m^v_{n\rightarrow m}=\frac{2r_n}{\sigma^2}+\sum_{m'\in\mathcal{M}(n)\setminus m}m^c_{m'\rightarrow n},
\end{eqnarray}
the total log-likelihood ratio (LLR) of $v_n$ is
\begin{eqnarray}\label{eqn:VLLR}
    L_n=\frac{2r_n}{\sigma^2}+\sum_{m\in\mathcal{M}(n)}m^c_{m\rightarrow n}.
\end{eqnarray}

\section{Adaptive Grouping and AGSBP Decoding}\label{section:Proposal}
\subsection{General Adaptive GSBP Decoder}\label{subsection:AGSBP}

Let the index set of the updated VNs (in an iteration) be denoted by $\mathcal{V}$ and its complement by
$\mathcal{V}^c=\mathcal{Z}_N\setminus \mathcal{V}$, where $\mathcal{Z}_N\triangleq \{0,1,\ldots,N-1\}$.
Suppose $l$ is the iteration counter and $l_{\max}$ is the maximum iteration number, then a generic AGSBP
decoding algorithm can be described by \textbf{Algorithm \ref{alg:ASGBP}}.

\begin{algorithm}[H]
    \caption{Adaptive Group Shuffled BP Algorithm}
    \label{alg:ASGBP}
    \textbf{Initialization} Set the iteration counter $l=0$.\\
    \textbf{Step 1} Set $\mathcal{V}^c=\mathcal{Z}_N$ and $\mathcal{V}=\mathcal{G}=\emptyset$. Let $l\leftarrow l+1$.\\
    \textbf{Step 2} Perform an adaptive grouping method (e.g.,
    \par\hskip\algorithmicindent\hskip\algorithmicindent
    \textbf{Algorithms} \textbf{\ref{alg:BS1}} or \textbf{\ref{alg:BS2}}) to form $\mathcal{G}$.\\
    \textbf{Step 3} Propagate $m^c_{m \rightarrow n}$ and then update $m^v_{n \rightarrow m} \forall m\in$
    \par\hskip\algorithmicindent\hskip\algorithmicindent
    $\mathcal{M}(n),~n\in\mathcal{G}$. \\
    \textbf{Step 4} Let $\mathcal{V}\leftarrow\mathcal{V}\cup\mathcal{G}$ and $\mathcal{V}^c\leftarrow\mathcal{V}^c\setminus\mathcal{G}$. If $\mathcal{V}^c=\emptyset$, go to
    \par\hskip\algorithmicindent\hskip\algorithmicindent
    \textbf{Step 5}; otherwise, go to \textbf{Step 2}.\\
    \textbf{Step 5} If a valid codeword is obtained or $l=l_{\max}$, stop
    \par\hskip\algorithmicindent\hskip\algorithmicindent
    decoding; otherwise, go to \textbf{Step 1}.
\end{algorithm}

Note that the VNs are grouped sequentially and the grouping is likely to vary in each iteration.
Once a new VN group is determined by \textbf{Step 2}, we update the C2V and V2C messages
associated with
the VNs belong to this group by \textbf{Step 3} before searching for the next VN group for updating.

\subsection{Adaptive VN Grouping Methods}\label{subsection:adap_full}
We now present integer-valued metrics for selecting VNs from $\mathcal{V}^c$. The selected VNs have
the least reliable tentative decisions and the highest probability of being corrected if updated.

Using the syndrome (checksum) vector $\bm{s}=(s_0,s_1,\cdots,$ $s_{M-1})=\hat{\bm{u}} \cdot\bm{H}^{T}$ (mod 2),
we define an integer-valued unreliability index
\begin{eqnarray}\label{eqn:BF}
    E_n=\Omega_n\left(\sum_{m\in\mathcal{M}(n)}s_m\right),
\end{eqnarray}
where
\begin{eqnarray}\label{eqn:mapping}
    \Omega_n(x)=\left\lfloor \frac{x}{d^v(n)}\cdot d^v_{\max}\right\rfloor
\end{eqnarray}
and $d^v_{\max}=\max_nd^v(n)$.
When $\mathcal{C}$ is a regular code, (\ref{eqn:BF}) becomes the flipping function of Gallager's BF decoding
algorithm \cite{LDPC} and is equal to the number of unsatisfied check nodes (UCNs) associated with VN $v_n$.
We further define
\begin{eqnarray}\label{eqn:Fn}
    F_n=\sum_{m\in\mathcal{M}(n)}q_{mn}s_m,
\end{eqnarray}
where
\begin{eqnarray}\label{eqn:q_mn}
    q_{mn}=\left\{
        \begin{array}{ll}
            1,&\quad \text{if}~\displaystyle\max_{n'\in\mathcal{N}(m)}E_{n'}=E_n~\text{and}~E_n\geq\eta\\
            0,&\quad \text{otherwise}\\
        \end{array}
    \right.
\end{eqnarray}
and $\eta$ is a numerically-optimized integer. (\ref{eqn:Fn}) counts the number of UCNs connected to VN $v_n$
for which it is a local $E_n$-maximizing VN. This function is similar to the reliability metric used in the
parallel weighted BF decoder \cite{PWBF}. As a bit decision associated with a large $F_n$ is likely to be
incorrect, we consider the VNs in $\mathcal{V}^c$ with the largest $F_n$ as the ones which have the least
reliable decision.

Let $\oplus$ denote the exclusive or (XOR) operation and $I^c_{m \rightarrow n}$ be the pre-computed sign bit
of the C2V message to be sent from $c_m$ to $v_n$, i.e., $I^c_{m \rightarrow n}\triangleq \BSGN(m^c_{m \rightarrow n})
=\bigoplus_{n'\in\mathcal{N}(m)\setminus n} \BSGN(m^v_{n' \rightarrow m})$, where $\BSGN(x)=\left[1-\text{sgn}(x)\right]/2$.
We need another integer-valued indicator
\begin{eqnarray}\label{eqn:An}
   A_n=\Omega_n\left(\sum_{m\in\mathcal{M}(n)} I^c_{m \rightarrow n}\oplus \BSGN(L_n)\right),
\end{eqnarray}
where the argument of $\Omega_n(\cdot)$ counts (predicts) the number of \emph{future} incoming messages
whose signs are different from that of the total LLR of $v_n$. The normalized count $A_n$ is then used to quantify
the force of driving a bit decision to change after updating, as a larger $A_n$ implies that the decision of $v_n$
may have a higher probability of being flipped. It is thus reasonable to conjecture that, among these local
$F_n$-maximizing VNs, the VNs which are most likely to be corrected after receiving related CN messages are the ones
which have the largest $A_n$. The VN selection procedure is summarized below.

\begin{algorithm}[H]
    \caption{Adaptive Grouping Method I}
    \label{alg:BS1}
    \textbf{Initialization} Set $\mathcal{G}=\emptyset$.\\
    \textbf{Step 1} $\forall n\in\mathcal{V}^c$, compute $F_n$, and find $F^*=\max_{n\in\mathcal{V}^c}F_n$.
    \par\hskip\algorithmicindent\hskip\algorithmicindent
    If $F^*=0$, stop and output $\mathcal{G}=\mathcal{V}^c$. \\
    \textbf{Step 2} Let $\mathcal{S}=\{n|F_n=F^*,n\in\mathcal{V}^c\}$ and compute $A_n$ for
    \par\hskip\algorithmicindent\hskip\algorithmicindent
    all $n\in\mathcal{S}$.\\
    \textbf{Step 3} Find $A^*=\max_{n\in\mathcal{S}} A_n$ and form the candidate set
    \par\hskip\algorithmicindent\hskip\algorithmicindent
    $\tilde{\mathcal{G}}=\{n|A_n=A^*,~n\in\mathcal{S}\}$ .\\
    \textbf{Step 4} Select $n^*$ arbitrarily from $\tilde{\mathcal{G}}$ and add $n^*$ to $\mathcal{G}$. Then,
    \par\hskip\algorithmicindent\hskip\algorithmicindent
    remove all $n \in \mathcal{N}(m),m\in$ $\mathcal{M}(n^*)$ from $\tilde{\mathcal{G}}$.\\
    \textbf{Step 5} If $\tilde{\mathcal{G}}=\emptyset$, stop and output $\mathcal{G}$; otherwise, go to \textbf{Step 4}.
\end{algorithm}
\squeezeup
In \textbf{Step 1}, we compute the $F_n$'s of those un-updated VNs. If all the resulting $F_n$'s are zero, i.e.,
$F^*=0$, we conclude that they are reliable and put these VNs in the same group. When $F^*>0$, we suspect that some
incorrect decisions may still exist in $\mathcal{V}^c$. Thus in \textbf{Step 2} and {\bf Step 3}, we collect the VNs
with largest $F_n$ value and select the ones having maximum $A_n$ to form a tentative set $\tilde{\mathcal{G}}$. In
\textbf{Step 4}, we randomly select one index from $\tilde{\mathcal{G}}$, say $n^*$, to join $\mathcal{G}$ and remove
it along with the indices of the VNs which are connected with those CNs linked to $v_{n^*}$ (i.e., $\mathcal{M}(n^*)$)
from $\tilde{\mathcal{G}}$. The purpose of removing these VNs is to prevent potential mutual erroneous message exchanges.
\textbf{Step 4} is repeated until $\tilde{\mathcal{G}}$ is emptied.

Since finding $\max_{n'\in\mathcal{N}(m)}E_{n'}$ for each UCN ($s_m=1$) in (\ref{eqn:q_mn}) requires extra computational
effort, a simple alternative is to use (\ref{eqn:BF}) as the reliability metric directly. Moreover, as the sign of the a
V2C message from a VN is likely to be the same as that of its LLR value in later iterations, we have
\begin{eqnarray}
    \nonumber
    I^c_{m \rightarrow n}
    &=&\bigoplus_{n'\in\mathcal{N}(m)\setminus n} \BSGN(m^v_{n' \rightarrow m})\\
    &\approx& \bigoplus_{n'\in\mathcal{N}(m)\setminus n} \BSGN(L_{n'})
\label{eqn:appr}
\end{eqnarray}
and therefore
\begin{eqnarray}\label{eqn:appr2}
    A_n \approx \Omega_n\left(\sum_{m\in\mathcal{M}(n)} \bigoplus_{n'\in\mathcal{N}(m)} \BSGN(L_{n'})\right)=E_n,
\end{eqnarray}
which indicates that the VNs with the largest $E_n$ may have the highest probability of being corrected as well.
By adopting (\ref{eqn:BF}) and (\ref{eqn:appr2}), we obtain a simplified version of \textbf{Algorithm \ref{alg:BS1}}.
\begin{algorithm}[H]
    \caption{Adaptive Grouping Method II}
    \label{alg:BS2}
    \textbf{Initialization} Set $\mathcal{G}=\emptyset$.\\
    \textbf{Step 1} Compute $E_n$ for all $n\in\mathcal{V}^c$.\\
    \textbf{Step 2} Let $E^*=\max_{n\in\mathcal{V}^c}E_n$. If $E^*<\delta$, stop and output
    \par\hskip\algorithmicindent\hskip\algorithmicindent
    $\mathcal{G}=\mathcal{V}^c$;
    otherwise, set $\tilde{\mathcal{G}}=\{n|E_n=E^*,n\in\mathcal{V}^c\}$.\\
    \textbf{Step 3} Select $n^*$ arbitrarily from $\tilde{\mathcal{G}}$ and add $n$ to $\mathcal{G}$. Then,
    \par\hskip\algorithmicindent\hskip\algorithmicindent
    remove all $n$ where $n \in \mathcal{N}(m),m\in$ $\mathcal{M}(n^*)$ from $\tilde{\mathcal{G}}$.\\
    \textbf{Step 4} If $\tilde{\mathcal{G}}=\emptyset$, stop and output $\mathcal{G}$; otherwise, go to \textbf{Step 3}.
\end{algorithm}

The integer reliability threshold $\delta$ in \textbf{Step 2} is numerically optimized through simulation.
No matter whether \textbf{Algorithms \ref{alg:BS1}} or \textbf{\ref{alg:BS2}} is used as the grouping method
in \textbf{Step 2} of the AGSBP algorithm (\textbf{Algorithm \ref{alg:ASGBP}}), the set of parameters, $\{A_n, F_n, E_n\}$
is immediately updated once any related message is renewed. The selection of next $\mathcal{G}$ is based on the
updated information and therefore the group partition and the corresponding group sizes for different iterations
may not be the same.

\section{Numerical Results}
\label{section:results}
We present the frame error rate (FER) performance of the conventional GSBP algorithm
and AGSBP algorithms with the proposed grouping methods in decoding MacKay's (1008, 504)
regular LDPC code (504.504.3.504 \cite{MacKayCode}, $d^v(n)=3$), (806, 272) regular code
(816.1A4.845 \cite{MacKayCode}, $d^v(n)=4$), and WiFi (1944, 972) quasi-cyclic (QC) LDPC
code \cite{WiFi}. The frame size is assumed to be equal to the codeword length, hence FER
is the same as the codeword error probability. AGSBP-I and AGSBP-II in the following figures
denote the AGSBP algorithms that use Adaptive Grouping Method I and II (\textbf{Algorithm
\ref{alg:BS1}} and \textbf{\ref{alg:BS2}}), respectively. For further decoding complexity
reduction, we also consider the min-sum (MS) approximation \cite{MSA} of the C2V updating
formula (\ref{eqn:C2V}). We denote the MS-based GS algorithms with the conventional
grouping method by GSMS. Similarly, the MS-based adaptive group shuffled decoders using
proposed grouping methods I and II are denoted by AGSMS-I and AGSMS-II, respectively. Table
\ref{TB:Parameters} lists the optimized parameters used for AGSBP and AGSMS algorithms in
decoding different codes.

\subsection{Numerical Examples}
\begin{table}[t]
\begin{center}
\caption{Parameter Values used for AGSBP/AGSMS Algorithms}
\label{TB:Parameters}
\scalebox{0.95}{
\begin{tabular}{|c|c|c|c|c|}\hline

\multirow{2}{*}{Code}& AGSBP-I & AGSBP-II  & AGSMS-I & AGSMS-II  \\\cline{2-5}
&$\eta$&$\delta$&$\eta$&$\delta$\\\hline
MacKay (1008,504) & 1 & 1 & 1 & 2\\\hline
MacKay  (816,272)& 1 &2 & 2 & 2\\\hline
WiFi (1944,972) & 1 & 4 & 6 & 6\\\hline

\end{tabular}
}
\end{center}
\end{table}

\begin{figure}[t]
\begin{center}
\epsfxsize=3.4in \epsffile{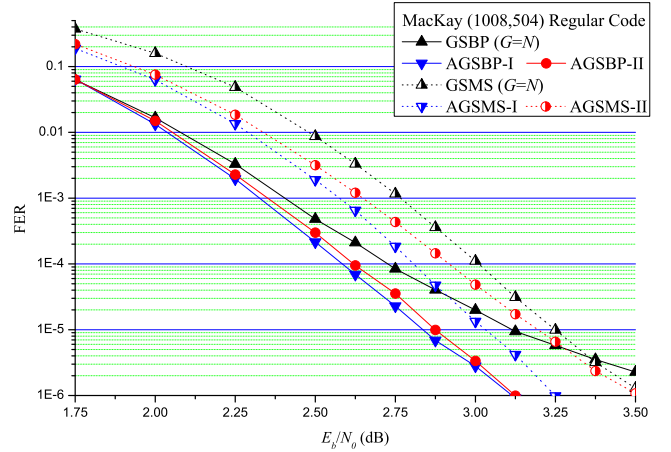}
\caption{\label{FER_M1K}
FER performance of Mackay's (1008, 504) regular LDPC code using various GS decoding algorithms.}
\squeezeup
\end{center}
\end{figure}

\begin{figure}[t]
\begin{center}
\epsfxsize=3.4in \epsffile{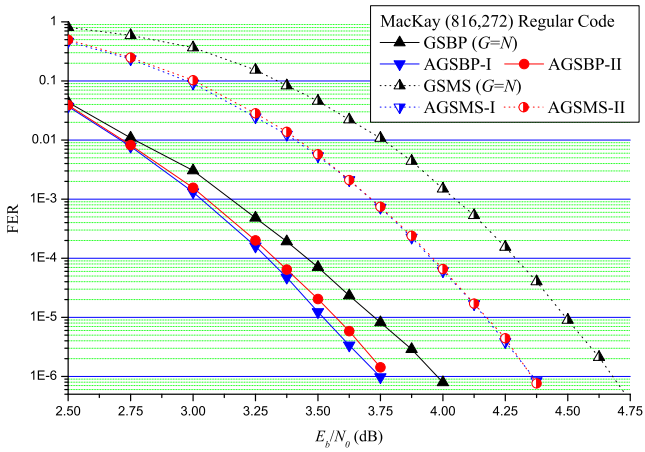}
\caption{\label{FER_M816}
FER performance of Mackay's (816, 272) LDPC code using various GS decoding algorithms.}
\squeezeup
\end{center}
\end{figure}

\begin{figure}[t]
\begin{center}
\epsfxsize=3.4in \epsffile{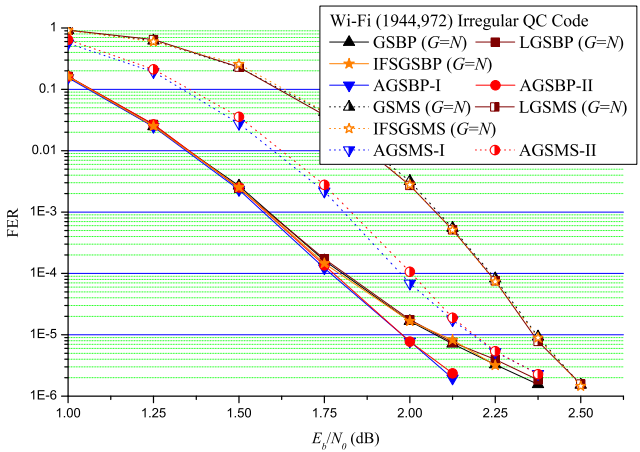}
\caption{\label{FER_W2K}
FER performance of IEEE 802.11 (1944, 972) rate-1/2 LDPC code using various GS decoding algorithms.}
\squeezeup
\end{center}
\end{figure}

Figs. \ref{FER_M1K} and \ref{FER_M816} respectively plot the FER performance of the (1008, 504)
and (806, 272) regular MacKay codes using various GS algorithms with $l_{\text{max}}=25$. For the
(1008, 504) code, the AGSBP-I and AGSBP-II algorithms achieve about 0.3 dB and 0.25 dB gains against
the conventional GSBP decoder at FER $\approx 10^{-5}$. The AGSMS algorithms outperform the GSMS
decoder as well. When decoding the (806, 272) code, the AGSBP-I algorithm gives a 0.25 dB gain
against the conventional GSBP decoder at FER $\approx10^{-5}$. The AGSBP-II algorithm outperforms
the GSBP one by 0.2 dB at the same FER. Furthermore, the decoding gains of both AGSMS-I/-II over
the GSMS decoder are about 0.3 dB at FER $\approx10^{-5}$.

Fig. \ref{FER_W2K} presents the FER performance of various algorithms in decoding the length-1944
WiFi code where $l_{\max}=50$. The performance of the local girth based GSBP (LGSBP) \cite{GSBP2},
informed fixed scheduling based GSBP (IFSGSBP) \cite{GSBP3} algorithms, and their MS-based variants
(denoted by LGSMS and IFSGSMS) is also shown in Fig. \ref{FER_W2K} for reference purpose. To limit
the implementation parallelism, we set the constraint for the WiFi code that the group size determined
by our grouping methods be less than 1944/3=648 VNs. Simulation results show that the AGSBP algorithms
yield about 0.25 dB performance gain over the GSBP, IFSGSBP, and LGSBP algorithms at the FER $\approx
2\times10^{-6}$. Moreover, by applying our grouping methods, AGSMS algorithms also offer improved
performance against the GSMS, IFSGSMS, and LGSMS algorithms.
\squeezeup
\subsection{Complexity Analysis}

All GSBP (GSMS) algorithms discussed, including our AGSBP (AGSMS) decoder, need
the same basic computing complexity. For our AGSBP and AGSMS algorithms,
extra computation is needed whenever \textbf{Steps 2} of \textbf{Algorithm
\ref{alg:ASGBP}} is activated.
$\Omega_n(x)$ can be obtained by using a look-up table since $d^v(n), d_{max}^v$ are known
and $x$ is an integer. The computation of $E_n$ can be done by having each UCN sending a
triggering signal to the counter associated with its connected VNs. The UCN number of an
ungrouped VN can then be accumulated (added), so the number of required additions is (at most)
$\sum_{m:s_m=1}d^c(m)$. The AGSBP-II and AGSMS-II decoders need to find $E^*$ which requires
$|\mathcal{V}^c|-1$ comparisons, where $|\cdot|$ denotes set cardinality. For the AGSBP-I
and AGSMS-I algorithms, besides $E_n$, they also have to compute $F_n, A_n$ and find $F^*$
and $A^*$. Given $E_n$, we need $\sum_{m:s_m=1}[d^c(m)-1]$ comparisons to find $\max_{n' \in
\mathcal{N}(m)}E_{n'}$ for all UCNs; see (\ref{eqn:q_mn}). As for $F_n$, it can be computed
in a way similar to that of computing $E_n$, and finding $F^*$ requires another $|\mathcal{V}^c|-1$
comparisons. To compute $A_n$ and find $A^*$, we need $\sum_{n\in\mathcal{S}}[d^v(n)-1]$
additions and $|\mathcal{S}|-1$ comparisons. Computing checksums and the XOR operations in
(\ref{eqn:An}) are omitted for they involve binary operations only.

Shown in Table \ref{TB:GSBP} is the basic computational complexity per iteration in decoding
MacKay's (1008, 504) code. The corresponding extra average per-iteration computational complexity
at some selected iterations for the proposed algorithms to decode the same code are listed in Table
\ref{TB:AGSBP}. The complexity is evaluated at specific SNRs and iterations. Note that the operations
listed in Table \ref{TB:GSBP} are real-number operations while those shown in Table \ref{TB:AGSBP}
are integer based; in fact, each integer addition is only a simple `add one' (accumulation) operation.

\begin{table}[t]
\begin{center}
\caption{Basic Complexity ($\times 10^3$) of GSBP and GSMS Algorithms}
\label{TB:GSBP}
\begin{tabular}{|c|c|c|c|}\hline
      & Addition/Subtraction & Comparison  & $\phi(\cdot)$-Operation\\\hline
GSBP  & 18.144               & 0           & 18.144           \\\hline
GSMS  & 6.048                & 12.096      & 0              \\\hline

\end{tabular}
\squeezeup
\end{center}
\end{table}

\begin{table}[t]
\begin{center}\caption{Simulated Average Extra Conditional Complexity ($\times 10^3$) for AGSBP
and AGSMS Decoders (AD: Addition; CP: Comparison)}
\label{TB:AGSBP}
\scalebox{0.95}{
\begin{tabular}{|c|c|c|c|c|c|c|c|c|c|}\hline

\multirow{2}{*}{SNR}& \multirow{2}{*}{Iter.} & \multicolumn{2}{c|}{AGSBP-I}  & \multicolumn{2}{c|}{AGSBP-II} & \multicolumn{2}{c|}{AGSMS-I}& \multicolumn{2}{c|}{AGSMS-II} \\\cline{3-10}

&&AD&CP&AD&CP&AD&CP&AD&CP\\\hline\hline

    & 5  &0.31 &5.16 &0.13 & 4.97 &0.65 &6.89  &0.19  &3.85  \\\cline{2-10}
2.75& 10 &1.85 &13.8 &0.95 & 14.6 &3.97 &21.2  &0.52  &4.87  \\\cline{2-10}
dB  & 15 &4.75 &26.0 &1.91 & 22.7 &7.04 &33.2  &1.13  &6.46  \\\cline{2-10}
    & 20 &4.71 &27.3 &1.90 & 24.1 &8.33 &37.3  &1.18  &6.82  \\\hline\hline

    & 5  &0.18 &4.13 &0.09 & 3.97 &0.35 &6.04  &0.12  &3.38  \\\cline{2-10}
3.0 & 10 &1.83 &13.3 &1.02 & 14.5 &3.34 &18.5  &0.42  &4.46  \\\cline{2-10}
dB  & 15 &3.71 &21.1 &1.92 & 22.4 &6.67 &31.0  &1.10  &6.34  \\\cline{2-10}
    & 20 &3.04 &19.6 &1.56 & 21.2 &7.16 &33.5  &1.29  &6.84  \\\hline

\end{tabular}
}
\squeezeup
\end{center}
\end{table}

\section{Conclusion}
\label{section:conclusion}
We have developed new VN grouping methods for use in GS decoding of LDPC codes. The proposed
methods employ integer based metrics to sequentially select the VNs for updating and can be
applied to both BP and MS based decoding algorithms. The extra binary/integer computational
efforts needed for the adaptive VN grouping methods are evaluated. We present some numerical
behaviors of the proposed AGSBP and AGSMS decoding algorithms and demonstrate that both of
them are able to provide improved performance in comparison with some known grouping methods
in decoding either regular or irregular LDPC codes.
\squeezeup

\end{document}